\documentclass[a4paper,11pt]{article}
\pdfoutput=1 % if your are submitting a pdflatex (i.e. if you have
\usepackage{jheppub} % for details on the use of the package, please
\usepackage[T1]{fontenc} % if needed

\usepackage{tikz-feynman}
\usepackage{subfig}
\usepackage{amsmath}
\usepackage{mathtools}
\usepackage{bm}
\usepackage{esvect}
\usepackage{verbatim}

\usepackage[utf8]{luainputenc}

\title{\boldmath 
Establishing a Search for $b \rightarrow s \ell^{+} \ell^{-}$ Anomalies at the LHC \\
}

 \author{Yoav Afik, Jonathan Cohen, Eitan Gozani, Enrique Kajomovitz, Yoram Rozen}
 \affiliation{Department of Physics, Technion: Israel Institute of Technology\\ Haifa, Israel}

\emailAdd{yoavafik@campus.technion.ac.il}
\emailAdd{jcohen@campus.technion.ac.il}
\emailAdd{gozani@campus.technion.ac.il}
\emailAdd{enrique@physics.technion.ac.il}
\emailAdd{rozen@physics.technion.ac.il}

\abstract{
One of the fundamental predictions of the Standard Model is Lepton Flavour Universality. Any deviation from this prediction would indicate the existence of physics beyond the Standard Model. Recent LHCb measurements present a pattern of deviations from this prediction in rare B-meson decays. While not yet statistically significant (currently $2.2-2.6 \sigma$), these measurements show an imbalance in the ratio of B-meson decays to a pair of muons  in association with a Kaon and decays to a pair of electrons in association with a Kaon. If the measured deviations are indeed present in nature, new physics may mediate interactions involving a pair of same flavour leptons, a $b$- and an $s$-quark. We present the prospect for a search of new physics in this type of interactions at the LHC, in a process that involves an $s$-quark, and a final state with two leptons and a $b$-jet. The proposed search can improve the sensitivity to new physics in these processes by a factor of four compared to current searches with in the total dataset expected at the LHC.

}

\begin{document} 
\maketitle
\flushbottom

\section{Introduction}
One of the fundamental predictions of the Standard Model (SM) is Lepton Flavour Universality (LFU). 
The photon and the $Z$-boson, independently, couple with the same strength to all three generations of leptons. 
%Similarly, the Z boson couples with a similar strength to all three generations of leptons. 
This property was tested extensively at LEP and SLD \cite{ALEPH:2005ab}. 
Any measured deviation from this prediction  implies the existence of new physics, beyond the SM (BSM).

Over the past several years, measurements of rare B-meson decays have exhibited deviations from the SM prediction \cite{Aaij:2014pli,Aaij:2017vbb,Aaij:2015esa,Aaij:2015oid,Wehle:2016yoi,Abdesselam:2016llu,ATLAS:2017dlm,CMS:2017ivg,Bifani:2017gyn}. These anomalies occur in the $b\to s \ell^+\ell^-$ transitions and are manifested by an imbalance in the ratio between the number of events measured in a final state with a pair of muons to a pair of electrons. 

The value of this ratio is stated within a given range of the lepton pair mass squared, $[q^2_{min}, q^2_{max}]$. The $q^2$-dependent partial width of the decay is $\Gamma$.
Currently there is a 2.6$\sigma$ discrepancy with respect to the SM \cite{Aaij:2014ora} in:
\begin{align}
  \begin{split}
&R_{K} = \frac{\int_{q^2_{min}}^{q^2_{max}} \frac{d\Gamma(B^{+} \to K^{+} \mu^+\mu^-)}{dq^2} dq^2}{\int_{q^2_{min}}^{q^2_{max}} \frac{d\Gamma(B^{+} \to K^{+} e^+e^-)}{dq^2} dq^2} \\%
&R_{K,[1.0,6.0]{\rm GeV^2}}=0.745^{+0.090}_{-0.074} \pm 0.036
  \end{split}
\label{eq:R_K}
\end{align} 

and a $2.1-2.5\sigma$ discrepancy in the related mode with the vector meson \cite{Aaij:2017vbb}:
\begin{align}
  \begin{split}
&R_{K^*} = \frac{\int_{q^2_{min}}^{q^2_{max}} \frac{d\Gamma(B^0 \to K^{0*} \mu^+\mu^-)}{dq^2} dq^2}{\int_{q^2_{min}}^{q^2_{max}} \frac{d\Gamma(B^0 \to K^{0*} e^+e^-)}{dq^2} dq^2} \\%
&R_{K^*,[0.045,1.1]{\rm GeV^2}}=0.66^{+0.11}_{-0.07} \pm 0.03 \\%
&R_{K^*,[1.1,6.0]{\rm GeV^2}}=0.69^{+0.11}_{-0.07} \pm 0.05
  \end{split}
\label{eq:R_K_star}
\end{align}

%Currently there is a 2.6$\sigma$ discrepancy with the SM in $R_K=\big(d\Gamma(B\to K\mu^+\mu^-)/dq^2\big)/\big(d\Gamma(B\to K e^+e^-)/dq^2\big)$ \cite{Aaij:2014ora},
%
%$$R_{K,[1,6]{\rm GeV^2}}=0.745\pm0.090$$
%
%and a $2.2-2.5\sigma$ discrepancy in the related mode with the vector meson, $R_{K^*}=\big(d\Gamma(B\to K^*\mu^+\mu^-)/dq^2\big)/\big(d\Gamma(B\to K^* e^+e^-)/dq^2\big)$ \cite{Bifani:2260258},
%
%$$R_{K^*,[0.045,1.1]{\rm GeV^2}}=0.66^{+0.11}_{-0.07}$$
%
%$$R_{K^*,[1.1,6]{\rm GeV^2}}=0.69^{+0.12}_{-0.08}$$
%
%
These anomalies imply that new physics may lay between the initial common state ($b$-quark) and the final state that is common between the measurements ($s$-quark and two charged leptons).
Many compelling UV-completions which aim to explain these anomalies include $Z'$ models, where the $b\to s \ell^+\ell^-$ transition is mediated at tree level by an exchange of a new heavy vector boson with flavour violating couplings to $b$- and $s$-quarks, as well as couplings to either electrons \cite{Carmona:2015ena} or muons~\cite{Kamenik:2017tnu,Descotes-Genon:2013wba,Buras:2013qja,Gauld:2013qja,Buras:2013dea,Altmannshofer:2014cfa,Celis:2015ara,Falkowski:2015zwa,Crivellin:2015lwa,Crivellin:2015mga,Boucenna:2016qad,Crivellin:2016ejn,Megias:2016bde,Megias:2017ove,GarciaGarcia:2016nvr,Allanach:2015gkd}.
The implications of such BSM models on proton-proton ($pp$) collisions have been examined by others ~\cite{Kohda:2018ZpI, Faroughy:2016osc}.
%(in the case of Ref.~\cite{Belanger:2015nma}  the latter is generated at loop level),  
Alternatively, the anomalies may also be explained via tree-level exchanges of lepto-quarks \cite{Hiller:2014yaa,Gripaios:2014tna,Becirevic:2015asa,Varzielas:2015iva,Fajfer:2015ycq,Becirevic:2016yqi}.
In order to keep our analysis as general as possible, here we consider an Effective Field Theory (EFT) approach with a four-point interaction between $b$-quark, $s$-quark, and a pair of charged leptons that can address the aforementioned anomalies.
%The other set of models generates the $b\to s \ell^+\ell^-$ through box loop diagrams with new heavy fields \cite{Gripaios:2015gra,Arnan:2016cpy}. 
 
%Therefore, in order to understand the kinematics of our "signal" we use an Effective Field Theory approach. 
%This will be discussed in more details at section \ref{sec:b_sll}.
%If new physics indeed lays in the aforementioned anomalies, it can be addressed by an Effective Field Theory (EFT) approach as a four-point interaction between $b$-quark, $s$-quark, and a pair of charged leptons.

The four point interaction may be meaningful also for processes that do not necessarily include B-meson decays. For instance, the interaction may play a role in the direct production of a $b$-quark and two opposite sign and same flavour leptons  in $pp$ collisions.
The correspondence of such an EFT model in these distinct energy scales has been previously studied \cite{Greljo:2017vvb}, demonstrating the applicability of measurements made in $pp$ collisions on the production mechanism on the physics of rare B-meson decays.
Taking into account the stated four-point interaction, an imbalance between the SM prediction and the recorded data in a final state with two same flavour leptons and one $b$-jet can be expected. Therefore, this kind of interaction can be tested using the extensive dataset recorded at the LHC.

We propose a search based on final states with exactly one $b$-jet and two opposite sign and same flavour leptons.
Any deviation from the SM prediction for this final state
%or for the ratio between muon and electron pairs, 
would imply that new physics lays in the interactions involved within this process. The proposed search differs significantly  from the analyses that target B-meson decays, mainly in the higher momentum thresholds of the objects involved in the final state. 
%It includes high $p_{T}$ leptons.
%One of the main advantages comparing to the B-meson analyses done at LHCb is the high trigger efficiency that can be achieved for electrons due to higher $p_{T}$ thresholds.
%Another advantage comparing to LHCb is that the amount of data recorded by ATLAS is significantly higher. 
%However, all of this do not promise the proposed analysis will have good sensitivity to test the anomalies observed by other experiments.
In the following sections we test the feasibility and estimate the sensitivity of the search at the LHC.

\label{intro}
\section{Theoretical Model}

%\subsection{Effective Field Theory for $b \rightarrow s \ell^{+} \ell^{-}$ Transitions}

%In order to have a model-independent analysis, we would like to take the most general approach that is possible.
%Therefore, a simple Effective Field Theory (EFT) will be used. 
In order to have a model-independent analysis, we consider an EFT setup including 6-dimensional operators.
Since the observed anomalies constitute a deviation in the number of muon pairs, only interactions between quarks and muons are considered (and not electrons).
The phenomena at low energies, i.e. a deficit in the number of muon pairs in B-meson decays, may be explained by an interference between the new physics contribution and the SM, that is smaller and opposite to the SM contribution.
However, here we consider higher energies and a parameter space where the resulting BSM contribution is significantly larger than the SM contribution that arise from electro-weak loop processes.
%in the parameter space we examine, the contribution from new physics is significantly larger than the SM contribution, which is via electro-weak loop processes.
%The SM contribution to similar final states via electro-weak loop processes was found to be negligible and was not taken into account.
We therefore assume a scenario where the LFU violation is dominated by the BSM contribution, and neglect the interference with the SM.
This leads to an enhancement in the number of expected muon pairs produced in association with a $b$-jet  at high di-muon masses due to BSM effects.

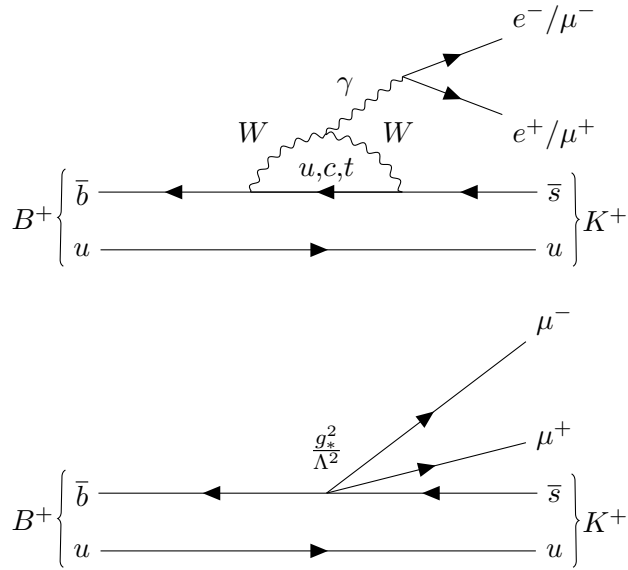
\begin{figure}
  \centering
  \subfloat{
\begin{tikzpicture}
%\label{fig:Feynman_bsll_SM}
  \begin{feynman}
    \vertex (a1) {\(\overline s\)};
    \vertex[left=2cm of a1] (a2);
    \vertex[left=2cm of a2] (a3);
    \vertex[left=2cm of a3] (a4) {\(\overline b\)};

    \vertex[below=2em of a4] (b1) {\(u\)}; 
    \vertex[below=2em of a1] (b2) {\(u\)}; 
    \vertex[above=2em of a4] (c1);
    \vertex[right=3.2cm of c1] (c2);
    \vertex[above=4em of a2] (d1);
    \vertex[above=4em of b2] (e1) {\(e^+ / \mu^+ \)};
    \vertex[above=4em of e1] (e2) {\(e^- / \mu^- \)};

    \diagram* {
      {[edges=fermion]
        (a1) -- (a2) -- (a3) -- (a4),
        (b1) -- (b2),
      },
            (a3) -- [boson, bend left ,edge label=\(W\)] (c2),
            (c2) -- [boson, bend left ,edge label=\(W\)] (a2),
            (c2) -- [boson ,edge label=\(\gamma\)] (d1),
            (a3) -- [edge label=\(u{,} c{,} t\)] (a2),
            (d1) -- [fermion] (e1),
            (d1) -- [fermion] (e2),

     };
    \draw [decoration={brace}, decorate] (b1.south west) -- (a4.north west)
          node [pos=0.5, left] {\(B^{+}\)};
    \draw [decoration={brace}, decorate] (a1.north east) -- (b2.south east)
          node [pos=0.5, right] {\(K^{+}\)};

%    \draw [decoration={brace}, decorate] (b1.south west) -- (a1.north west)
%          node [pos=0.5, left] {\(B^{0}\)};
%    \draw [decoration={brace}, decorate] (c3.north east) -- (c1.south east)
%          node [pos=0.5, right] {\(\pi^{-}\)};
%    \draw [decoration={brace}, decorate] (a6.north east) -- (b5.south east)
%          node [pos=0.5, right] {\(\pi^{+}\)};

  \end{feynman}
\end{tikzpicture}
% \caption{A representative Feynman diagram for a decay of a $B^{+}$ meson to a $K^{+}$ meson in association with two leptons in the SM.}
%\end{figure}
}

%\begin{figure}
%  \centering
  \subfloat{
\begin{tikzpicture}
  \begin{feynman}
    \vertex (a1) {\(\overline s\)};
    \vertex[left=3cm of a1] (a2);
    \vertex[above=0.5em of a2] (l2) {\(\frac{g_{*}^2}{\Lambda^{2}}\)};
    \vertex[left=3cm of a2] (a3) {\(\overline b\)};

    \vertex[below=2em of a3] (b1) {\(u\)}; 
    \vertex[below=2em of a1] (b2) {\(u\)}; 
    \vertex[above=4em of a2] (d1) ;
    \vertex[above=4em of b2] (e1) {\(\mu^+\)};
    \vertex[above=4em of e1] (e2) {\(\mu^-\)};

    \diagram* {
      {[edges=fermion]
        (a1) -- (a2) -- (a3),
        (b1) -- (b2),
      },
            (a2) -- [fermion] (e1),
            (a2) -- [fermion] (e2),

     };
    \draw [decoration={brace}, decorate] (b1.south west) -- (a3.north west)
          node [pos=0.5, left] {\(B^{+}\)};
    \draw [decoration={brace}, decorate] (a1.north east) -- (b2.south east)
          node [pos=0.5, right] {\(K^{+}\)};

%    \draw [decoration={brace}, decorate] (b1.south west) -- (a1.north west)
%          node [pos=0.5, left] {\(B^{0}\)};
%    \draw [decoration={brace}, decorate] (c3.north east) -- (c1.south east)
%          node [pos=0.5, right] {\(\pi^{-}\)};
%    \draw [decoration={brace}, decorate] (a6.north east) -- (b5.south east)
%          node [pos=0.5, right] {\(\pi^{+}\)};

  \end{feynman}
\end{tikzpicture}
}

 \caption{Representative Feynman diagrams for a decay of a $B^{+}$ meson to a $K^{+}$ meson in association with two leptons in the SM (upper) and in the EFT described in the text (bottom).
 Only muons are considered for the decay within the EFT approach.}
 \label{fig:Feynman_bsll_B}
\end{figure}

\begin{figure}
  \centering
\begin{tikzpicture}
\label{fig:Feynman_bsll}
  \begin{feynman}
    \vertex (a1) {g};
    \vertex[right=4cm of a1] (a2);
    \vertex[right=4cm of a2] (a3) {\(s\)};
    \vertex[below=12em of a3] (b1) {\(\overline b\)};
    \vertex[left=4cm of b1] (b2);
    \vertex[left=4cm of b2] (b3) {g};
    \vertex[below=6em of a2] (c1);
    \vertex[right=1cm of c1] (c2);
    \vertex[below=1em of a2] (d1);
    \vertex[above=1em of b2] (d2);
    \vertex[right=2cm of c2] (e1);
    \vertex[right=1cm of e1] (e2);
    \vertex[above=1em of e2] (e3) {\(\mu^+ \)};;
    \vertex[below=1em of e2] (e4) {\(\mu^- \)};
    \vertex[left=0.1cm of c2] (l2) {\(\frac{g_{*}^{2}}{\Lambda^{2}}\)};

    \diagram* {
      {[edges=fermion]
        (d1) -- (a3),
        (b1) -- (d2),
 %       (d1) -- (d2),
      },
      (a1) -- [gluon] (d1),
%      (b1) -- [fermion] (d2),
%      (d1) -- [fermion] (a3),
%      (b1) -- [fermion] (d1),
      (d2) -- [gluon] (b3),
      (d1) -- [anti fermion, edge label=\(s\)] (c2),
      (d2) -- [fermion, edge label=\(\overline b\)] (c2),
%      (c2) -- [scalar, edge label=\(A / a\)] (e1),
      (c2) -- [anti fermion] (e3),
      (c2) -- [fermion] (e4),
%      (c2) -- [edge label=\(\phi / a\)] (e1),

    };

  \end{feynman}
\end{tikzpicture}
	    \caption{A representative Feynman diagram for a production of one $b$-jet in association with two muons within the EFT approach.}
  \label{fig:Feynman_bsll_ATLAS}
\end{figure}
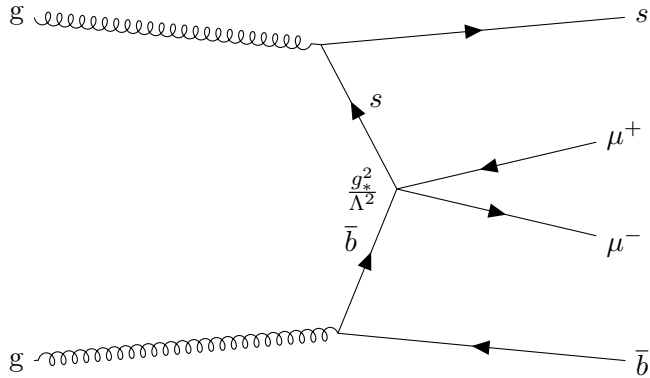

Representative Feynman diagrams for the decay of a B-meson within the scope of the SM and within the EFT approach are shown at Figure \ref{fig:Feynman_bsll_B}.
A representative Feynman diagram for tree level production of this model at the LHC is shown at Figure \ref{fig:Feynman_bsll_ATLAS}.

The effective Lagrangian of the benchmark model, as stated at \cite{Greljo:2017vvb}, is:
\begin{align}
\mathcal{L}_{eff} = \frac{C_{ij}^{U \mu}}{v^2} (\bar{u}_{L}^{i} \gamma_{\mu} u_{L}^{j}) (\bar{\mu}_{L} \gamma_{\mu} \mu_{L}) + \frac{C_{ij}^{D \mu}}{v^2} (\bar{d}_{L}^{i} \gamma_{\mu} d_{L}^{j}) (\bar{\mu}_{L} \gamma_{\mu} \mu_{L}),
\label{eq:Lagrangian_bsll}
\end{align}

where $C_{ij}^{U \mu}$ and $C_{ij}^{D \mu}$ are matrices that carry the flavour structure of the operators.
% $u_{L}^{i}$ and $d_{L}^{i}$ are left-handed up- and down-type singlets, $v$ is the VEV corresponding to a value of $246 GeV$, $\mu_{L}$ are left handed muons and $\gamma_{\mu}$ are the gamma matrices.
For the off-diagonal elements only the $b-s$ admixtures are considered, since those are the ones related to the observed anomalies.
The matrices take the form:
\begin{align}
C_{ij}^{U \mu} = 
\begin{pmatrix} C_{u \mu} & 0 & 0 \\ 0 & C_{c \mu} & 0 \\ 0 & 0 & C_{t \mu} \end{pmatrix} 
,
C_{ij}^{D \mu} = 
\begin{pmatrix} C_{d \mu} & 0 & 0 \\ 0 & C_{s \mu} & C_{b s \mu}^{*} \\ 0 & C_{b s \mu} & C_{t \mu} \end{pmatrix} 
\label{eq:Matrices_bsll}
\end{align} 

The generality of eq. \ref{eq:Lagrangian_bsll} stems from the fact that it can be accommadated by a plethora of new physics scenarios. Comprehensive EFT studies including the operators in eq. \ref{eq:Lagrangian_bsll} have recently been performed in the context of B-meson decays in \cite{Celis:2017gii,Cornella:2018tfd}. The Wilson coefficients can then be matched to specific model realizations, including $Z'$  and leptoquarks \cite{Faisel:2017glo,Fox:2018ldq,Allanach:2017bta,Fuyuto:2017sys,Alok:2017sui,Alok:2017jaf,Altmannshofer:2016jzy,Becirevic:2017jtw}, Composite Higgs \cite{Carmona:2017fsn}, Randall-Sundrum models \cite{DAmbrosio:2017wis,Blanke:2018sro}, Type III 2HDM \cite{Arhrib:2017yby}, $R$-parity violating supersymmetry \cite{Altmannshofer:2017poe} and more. 

%Matching at the tree level this operator to the standard effective weak Hamiltonian describing $b \to s$ transitions,

As for the $b \rightarrow s \ell^{+} \ell^{-}$ transitions, the recent combined fit reported \cite{Capdevila:2017bsm}:
\begin{align}
	\frac{\pi}{\alpha V_{t b} V^*_{t s}} C_{bs\mu} = -0.62 \pm 0.13~
	\label{eq:DC9}
\end{align}

Where $\alpha$ is the electromagnetic fine structure constant. The presumed scale of the new physics, $\Lambda$, is estimated by defining
\begin{align}
	C_{bs\mu} = g_{*}^2 v^2 / \Lambda^2.
	\label{eq:OpNorm}
\end{align}  
Taking the CKM matrix elements to be $|V_{ts}| = (40.0 \pm 2.7) \times 10^{-3}$ and $|V_{tb}| = 1.009 \pm 0.031$ \cite{Patrignani:2016xqp},
and using result \ref{eq:DC9},  the scale of new physics is: $\Lambda/g_{*} \approx 31^{+4}_{-3}$ TeV.

The present 95\% CL limit from the $13$ TeV ATLAS $pp \to \mu^+\mu^-$ analysis with 36 fb $^{-1}$ \cite{ATLAS:2017wce} and a prediction  for 3000 fb$^{-1}$ of luminosity as stated in \cite{Greljo:2017vvb} is:
\begin{align}
	\left| \frac{\pi}{\alpha V_{t b} V^*_{t s}} C_{b s\mu} \right| < 100 (39),
	\label{eq:Cbslimit}
\end{align} 

which corresponds to $\Lambda/g_{*} > 2.5 (3.9)$ TeV.
The stated ATLAS analysis, however, is aimed for final states with a pair of electrons or a pair of muons.
%contact interaction searches of 4-fermion operators of the shape $q\bar{q} \ell^{+} \ell^{-}$.
In contrast to the previous selection method, which is inclusive for a pair of electrons or a pair of muons, we 
add requirements on a few more variables, including the number of $b$-tagged jets.
%the invariant mass of the muon pair and the missing transverse energy.
As a result, the background is being reduced significantly.
%Many other observables can reduce the background significantly, while keeping most of the signal.
Therefore, a selection dedicated to the presented model is expected to improve the sensitivity significantly.

\label{model}
\section{Analysis}

\subsection{Simulated Event Samples}

Monte Carlo (MC) simulated event samples of $pp$ collisions at $\sqrt{s} = 13$ TeV were used to estimate the
background SM processes as well as the EFT signal. The SM processes
taken into consideration are: top pair production events ($t\overline{t}$);
top pair production events in association with an electro-weak boson ($t\overline{t}+W/Z$);
production of $Z$ boson and Drell-Yan processes in association with jets ($Z/\gamma^{*}+jets$);
and di-boson production ($WZ,WW$) processes. The background samples were uniformly
generated at leading order using {\sc MadGraph5\_aMC@NLO 2.6.1} \cite{Alwall:2011mad}  in the 5 flavour scheme
with the NNPDF30LO PDF set and interfaced with the {\sc Pythia 8.23} \cite{Mrenna:2016pyt} parton
shower. The {\sc MadSpin} interface was used to model the decay of top quarks
and the electro-weak bosons. 

A valid {\sc MadGraph} model was built in order to generate signal events, using the model presented in section \ref{model} as a benchmark. 
The signal samples were uniformly
generated at leading order using {\sc MadGraph5\_aMC@NLO 2.6.1} \cite{Alwall:2011mad}  in the 4 flavour scheme\footnote{We find this choice suitable since the appearance of a massive $b$-quark in the final state is indicative of the signal.}.
The signal samples were then normalized according to the total cross-section obtained in the 5 flavour scheme. 

Finally, all simulated samples were processed through
{\sc DELPHES 3} \cite{Favereau:2013del}  in order to simulate the detector effects and apply simplified
reconstruction algorithms.

%Interference effects between the signal model and the SM was found, in leading order and the 4 flavour scheme, to be negligible.
%The signal was generated at leading order using the 4-Flavour Scheme (4FS). 

\subsection{Event Reconstruction}

The analysis relies on the reconstruction of hadronic jets, muons,
and missing transverse energy ($E_{\mathrm{T}} ^ {\mathrm{miss}} $). Jets were reconstructed
using the anti-$k_{t}$~\cite{Cacciari:2008gp} clustering algorithm with a radius parameter
$R=0.4$ implemented in FastJet~\cite{Cacciari:2011ma,hep-ph/0512210}, and are required to have transverse momentum $p _ {\mathrm{T} } >20$ GeV
and pseudo-rapidity $\left|\eta\right|<2.5$. The identification of $b$-tagged jets
was done by applying a $p_ {\mathrm{T}}$-dependent weight based on the jet's
associated flavour, using truth information, and the MV2c20 tagging
algorithm~\cite{ATL-PHYS-PUB-2015-022} in the 70\% working point. Muons are reconstructed using truth-level
muons after applying an artifical $p_{\mathrm{T}}$- and $\eta$-dependent
efficiency weight, with isolation requirements from other energy-flow
objects applied in a cone of $R=0.5$, and a minimum $p_{\mathrm{T}}$ requirement of $20$ GeV for each muon.
The missing transverse energy
was calculated based on all energy-flow objets in the particle flow
approach.

\subsection{Event Selection}

An optimization was done by maximizing the sensitivity of the selection. 
%The term for the sensitivity mentioned in section \ref{sec:DM_simp}, (see eq. \ref{eq:Sensitivity_DM}), was not used since it does not treat well cases of very low background yield.
The sensitivity was estimated as the expected $Z$-value using the
\verb|BinomialExpZ| function by \verb|RooFit| \cite{Verkerke:2003ir}.
%\begin{align}
%Sensitivity = \frac{H_{1} - H_{0}}{{unc(H_{0})}} 
%\label{eq:Sensitivity_bsll}
%\end{align} 
%Assuming $H_{1} = S_{\mu \mu} + B_{\mu \mu}$, $H_{0} = B_{\mu \mu}$, $unc(H_{0}) = \sqrt{B_{\mu \mu} + (\sigma_{B_{\mu \mu}} \cdot B_{\mu \mu})^2}$,
%$$Sensitivity = \frac{S_{\mu \mu}}{\sqrt{B_{\mu \mu} + (\sigma_{B_{\mu \mu}} B_{\mu \mu})}} $$
%where $S_{\mu \mu}$ is the signal yield for muon pairs, $B_{\mu \mu}$ is the background yield for muon pairs
%and $\sigma_{B_{\mu \mu}}$ is the relative systematic uncertainty on the background. 
Different scenarios for the total relative uncertainty on the background were tested: $25\%, 50\%, 100\% $.
The integrated luminosity was chosen to be 120 fb$^{-1}$, which is an evaluation for the full Run-2 integrated luminosity, expected to be available by the end of 2018.

\begin{figure}
\centering
\subfloat{
\includegraphics[width=.50\textwidth]{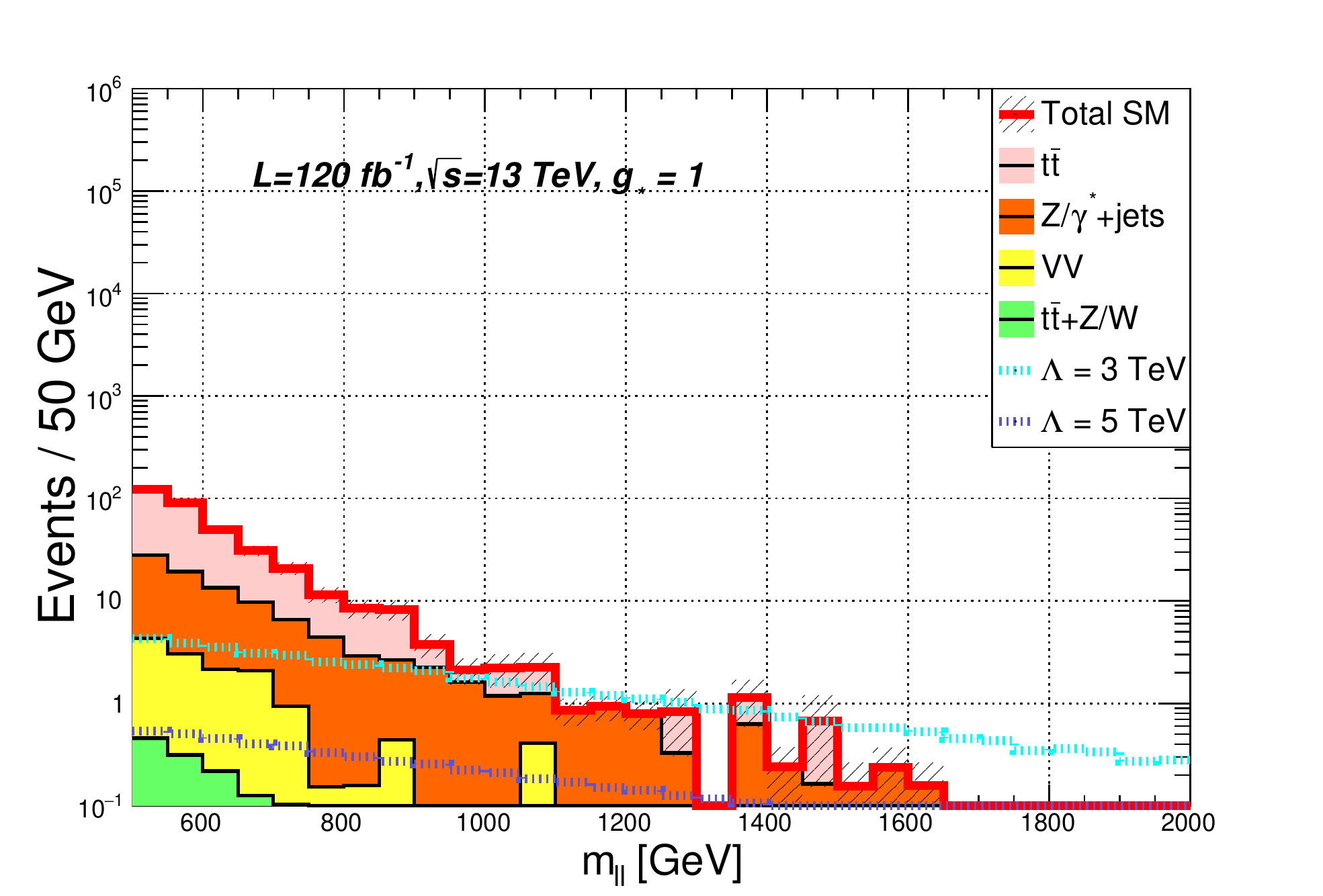}
%\caption{}
}
\subfloat{
\includegraphics[width=.50\textwidth]{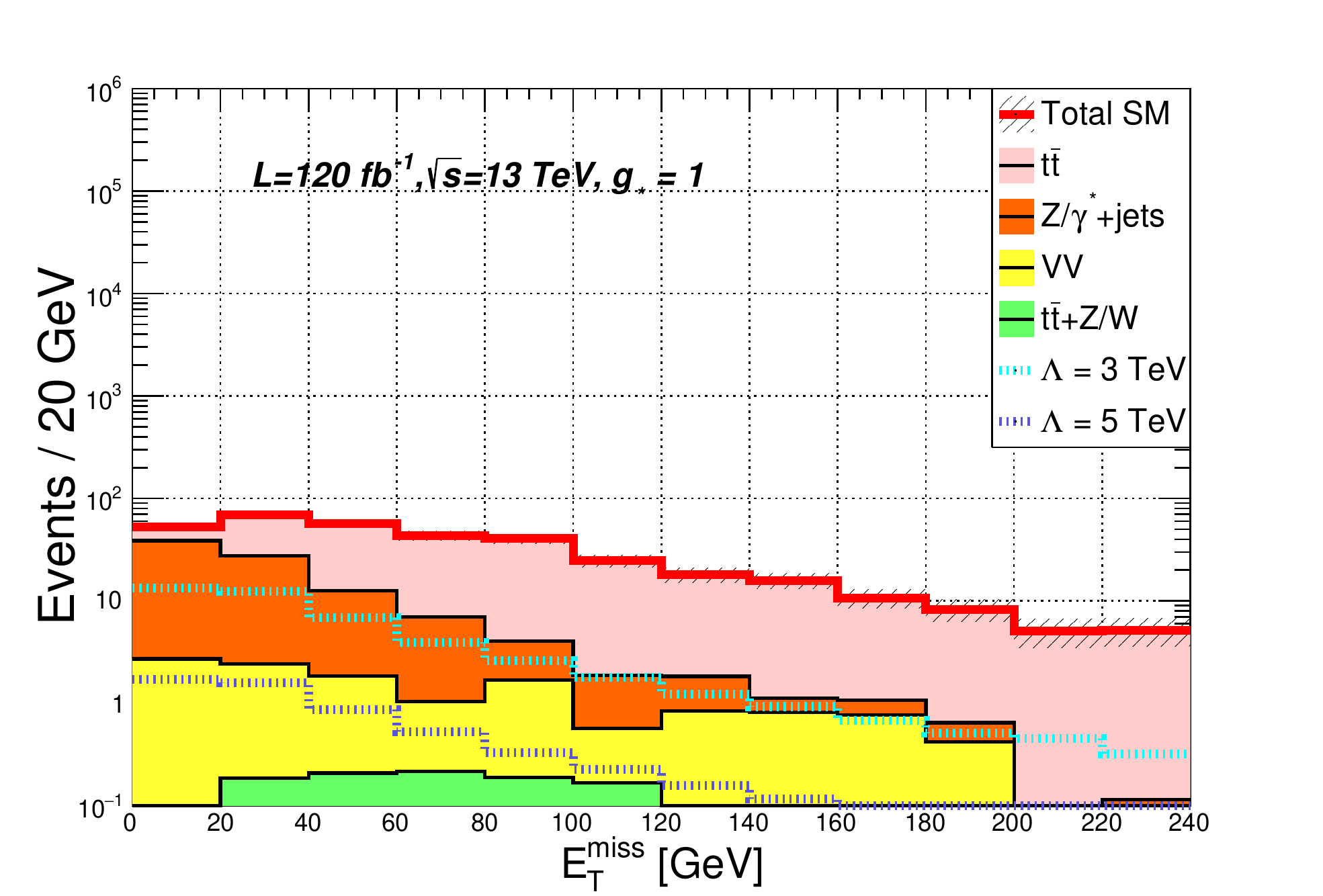}
%\caption{}
}\\
%\subfloat{
%\includegraphics[width=.50\textwidth]{bsll/leading_lep_pt_mumu_b.pdf}
%%\caption{}
%}
%\subfloat{
%\includegraphics[width=.50\textwidth]{bsll/sub_leading_lep_pt_mumu_b.pdf}
%%\caption{}
%}\\
%
%\caption{Distribution of the invariant mass of both muons (upper left), $E_{T}^{miss}$ (upper right) and transverse momentum of leading (lower left) and sub-leading (lower right) $p_{T}$ muon. The selection of two opposite sign muons and one $b$-jet are applied. The signal scenarios of $\Lambda = (1,3,5,10) TeV$ are presented as well.
%}
\caption{Distribution of the invariant mass of both muons $m_{\mu \mu}$ (left) and the missing transverse energy $E_{\mathrm{T}} ^ {\mathrm{miss}}$ (right). The selection of two opposite sign muons and exactly one $b$-jet are applied. The signal scenarios of $\Lambda = (3,5)$ TeV are presented as well, using $g_{*} = 1$. The requirement on the invariant mass of both muons $m_{\mu \mu} > 500$ GeV is applied. 
}
\centering
\label{fig:bsll_observables}
\end{figure}

As a base-point, the selection contains two muons with opposite-sign charges (OS).
In addition, since the signal is expected to have exactly one $b$-jet,
this selection was applied as well, using the medium working point of 70\% ($\mathcal{N}_{b}$).
Since the signal expected to have low missing transverse energy ($E_{\mathrm{T}} ^ {\mathrm{miss}} $), an upper selection on $E_{\mathrm{T}} ^ {\mathrm{miss}}$ was optimized. 
This is especially helpful in reducing the $t\overline{t}$ background, since in addition to the muons it contains two neutrinos that escape the detectors.
The invariant mass of both muons ($m_{\mu \mu}$) was used for optimization where new physics is expected to enter at the tail of the $m_{\mu \mu}$ distribution while keeping a small yield for the background.
%To reduce the background from events containing muons originating from the decay of a $Z$ boson in association with jets, events containing muons with invariant mass smaller within a band of $20 GeV$ from the mass of the $Z$ (91 GeV) were rejected.
%Two more variables that were used for the optimization are the leading $p_{T}$ ($p_{T}(\mu_{1})$) and sub-leading $p_{T}$ ($p_{T}(\mu_{2})$) muons.
The selection applied is summarized in table \ref{tab:bsll_mumu_selections}.
The distributions of the observables described in the text, within the selection of the muon pair and the b-tagged jets, are presented in figure \ref{fig:bsll_observables}.

\begin{figure}
\centering
\includegraphics[width=.70\textwidth]{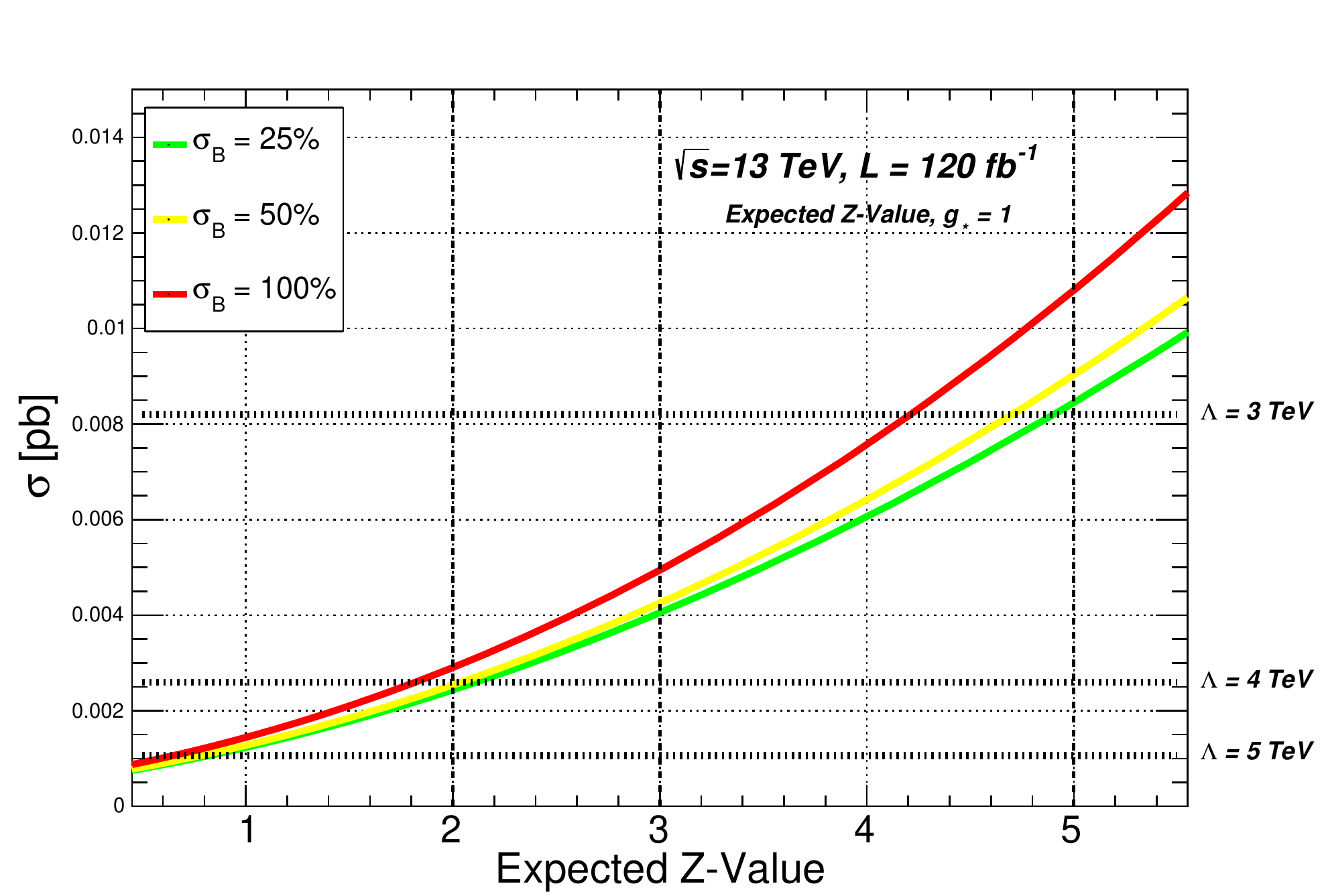}

\caption{The signal cross-section as a function of the expected $Z$-value, using the selections presented at table \ref{tab:bsll_mumu_selections}.
The cross-sections corresponding to $\Lambda=(3,4,5)$ TeV are presented, using $g_{*}=1$.}
\centering
\label{fig:Z_xsec}
\end{figure}

\begin{figure}
\centering
\includegraphics[width=.70\textwidth]{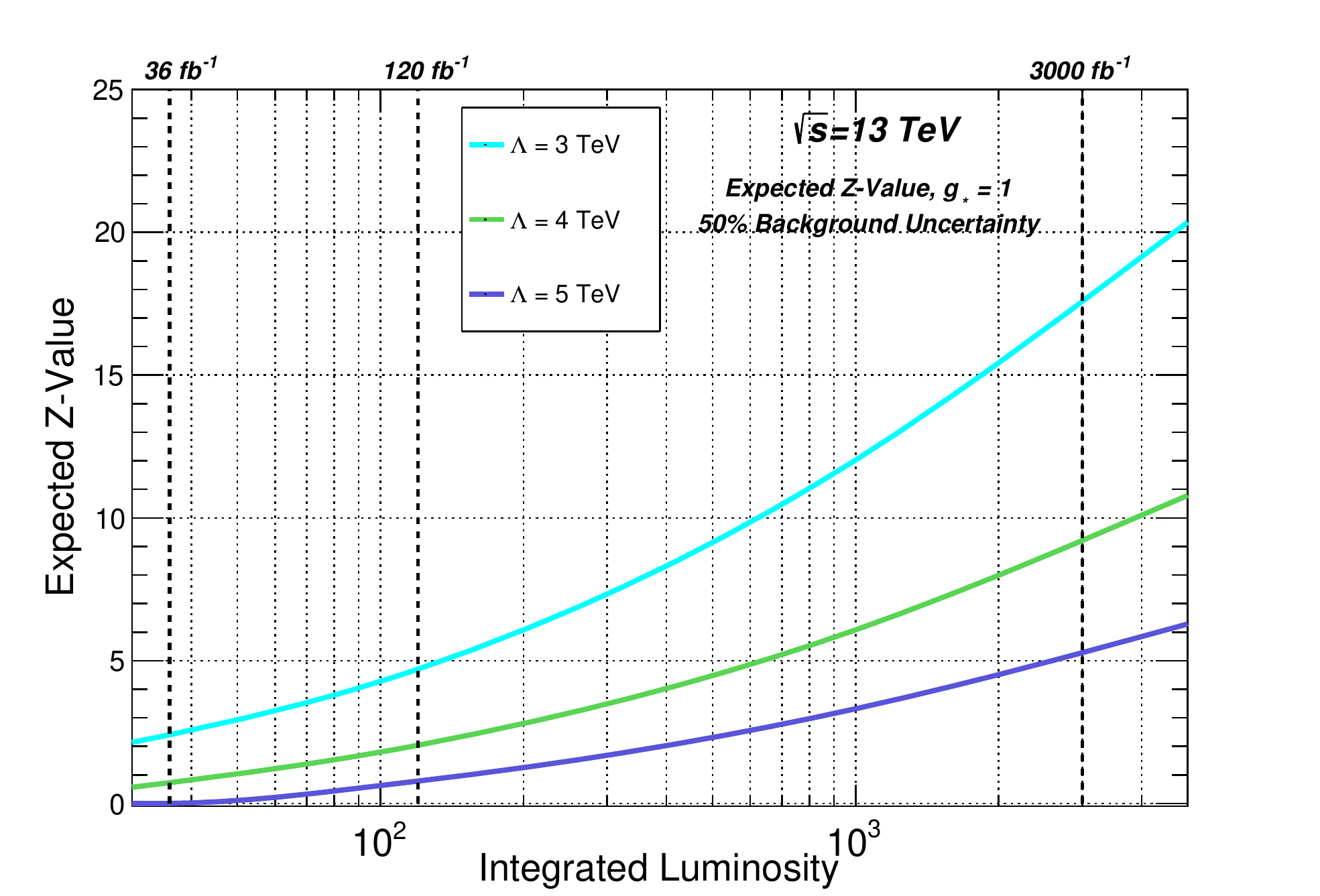}

\caption{The expected Z-Value as a function of the total integrated luminosity, using the selections presented at table \ref{tab:bsll_mumu_selections}.
The Z-Values corresponding to $\Lambda=(3,4,5)$ TeV are presented, using $g_{*}=1$.
}
\centering
\label{fig:Z_luminosity}
\end{figure}

The expected $Z$-value for the total integrated luminosity of 120 fb$^{-1}$, using the selections presented at table \ref{tab:bsll_mumu_selections}, is shown at Figure \ref{fig:Z_xsec}.
The expected $Z$-value as a function of the total integrated luminosity, for different parameter choice of the signal, using the selections presented at table \ref{tab:bsll_mumu_selections}, is shown at Figure \ref{fig:Z_luminosity}.

%The variables used for optimization are the missing transverse energy ($E_{T}^{miss}$), the invariant mass of both muons ($m_{\ell \ell}$) and the transverse momentum of both muons (leading and sub-leading lepton $p_{T}$). 

%The best combination of the cuts is:
%\begin{itemize}
%  \item[-] $\mathcal{N}_{b} = 1$
%  \item[-] $E_{T}^{miss} < 100 GeV$ 
%  \item[-] $m_{\ell \ell} > 100 GeV$
%  \item[-] Leading lepton $p_{T} > 200 GeV$
%  \item[-] sub-Leading lepton $p_{T} > 200 GeV$
%\end{itemize}

%//////////
%Another approach that was used for optimization is using the ratio between muon and electron pairs.
%Assuming the new physics correspond to a new interaction only with muons, $H_{1} = \frac{S_{\mu \mu} + B_{\mu \mu}}{B_{ee}}$, $H_{0} = \frac{B_{\mu \mu}}{B_{ee}}$,
%$unc(H_{0}) = \sqrt{(\frac{\sigma_{B_{\mu \mu}}}{B_{ee}})^{2} + (\frac{B_{\mu \mu} \sigma_{B_{ee}}}{B_{ee}^{2}})^{2}} $
%
%optimization was done by maximizing the term:
%$$Sensitivity = \frac{H_{1} - H_{0}}{{unc(H_{0})}} $$
%The $H_{1}$

\begin{table}[htp]
\begin{center}
%\resizebox{\textwidth}{!}{ 
\begin{tabular}{| l | c |}
\hline
Observable & Selection   \\
\hline
$\mathcal{N}_{b}$ & =1 \\
$\mathcal{N}_{\mu}$ & = 2, OS \\
$E_{T}^{miss}$ [GeV] & $< 180$ \\
$m_{\mu \mu}$ [GeV] & $> 1700$ \\
%$p_{T}(\mu_1)$ & $> 700$ \\
%$p_{T}(\mu_2)$ & $> 400$ \\
\hline
\end{tabular}
\end{center}
  \caption{Summary of the selection for muon pairs. The object definition and common selection described in the text are applied.}
\label{tab:bsll_mumu_selections}
\end{table}

\

\label{analysis}
\section{Ratio Analysis}

Keeping in mind the LHCb measurements \cite{Aaij:2014pli,Aaij:2014ora,Aaij:2015esa,Aaij:2015oid,Wehle:2016yoi,Abdesselam:2016llu,ATLAS:2017dlm,CMS:2017ivg,Bifani:2017gyn}, which hinted a deficit in the number of muon pairs with respect to the number of electron pairs, we suggest a ratio method as a complementary analysis.
While an increase in the number of muon pairs can be observed over the expected background if the increase is of certain significance, a deficit can only be observed if the background is well known and the deficit is larger than the background resolution (this statement is also true for an increase but an increase is in principle unlimited). Hence, in order for our model  to also accommodate a deficit at the higher muon pair invariant mass regime, a ratio method is proposed. This method has the advantage that the SM background does not need to be known to high precision as the systematical uncertainties associated with it vanish. 
Here too, we depart from the LHCb exclusive analysis and target the high invariant mass of the lepton pairs regime in the presence of a b-jet.

An evaluation of the signal cross-section as a function of the sensitivity was done, by using the selections presented at table \ref{tab:bsll_mumu_selections}.
Figure \ref{fig:Z_xsec} presents the signal cross-section as a function of the expected corresponding $Z$-value.
The expected $Z$-value was calculated using different relative uncertainties assumptions on the background.
In general, small differences observed between different values of the relative background uncertainty.
For this reason, looking at the ratio between final states with a muon pair and with an electron pair will not improve our sensitivity.
On the other hand, if we decide to look only at muon pairs in the final state, we can use similar selection for electron pairs as a CR, in order to normalise the expected SM yield.
%Figure \ref{fig:Z_xsec} presents the difference in the sensitivity for relative systematic uncertainties of 0\%, 10\%, 20\%, 30\%.
%The differences are very small, leading to the conclusion that the alternative analysis will not reduce the relative uncertainty.

\label{ratio}
\section{Results}
%\subsection{Reach at end of Run-2}

\begin{figure}
\centering
\includegraphics[width=.70\textwidth]{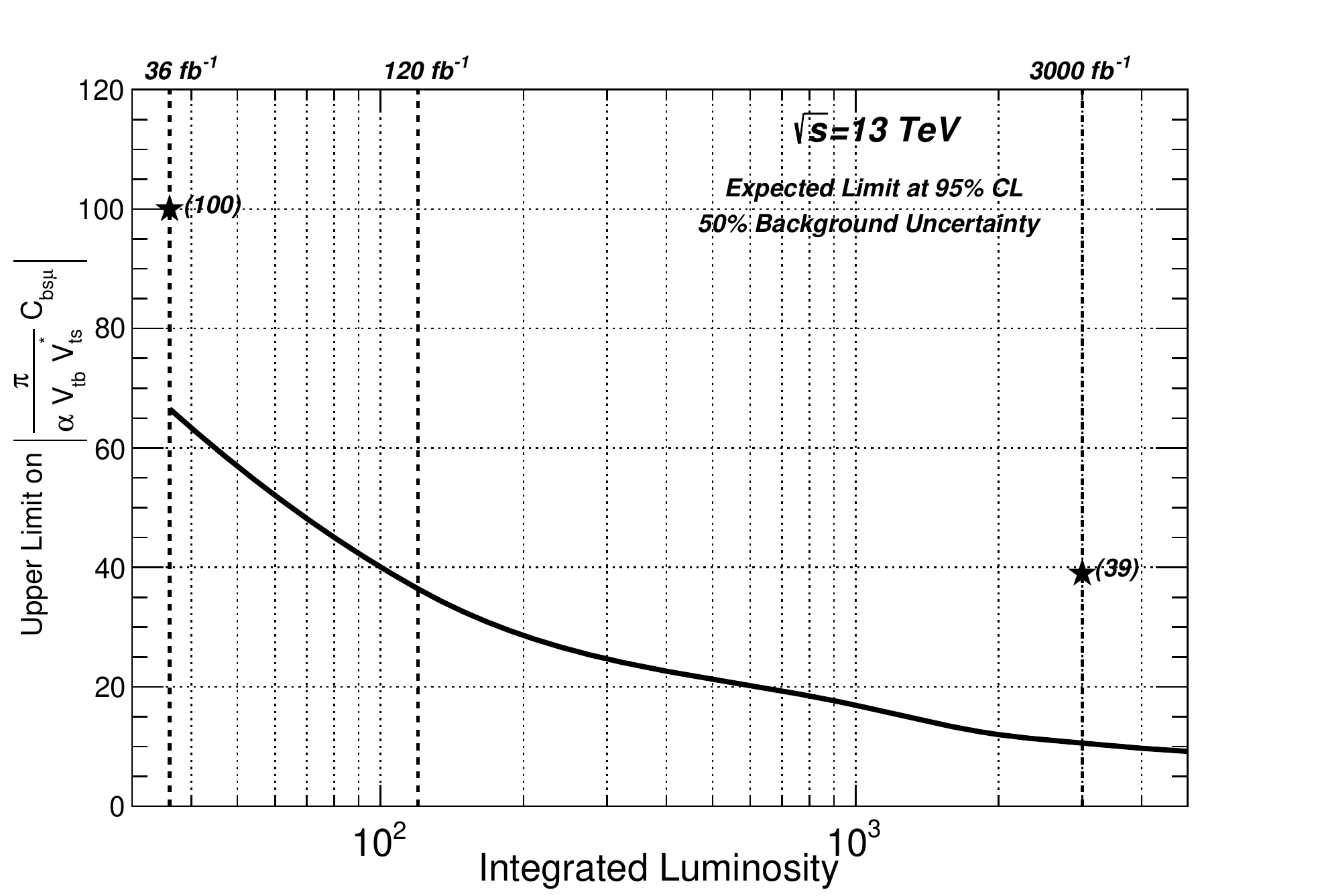}

\caption{Predicted upper limit on the term presented at equation \ref{eq:Cbslimit}, as a function of the total integrated luminosity.
The dashed lines correspond to integrated luminosity of (36, 120, 3000) fb$^{-1}$.
The upper limit for integrated luminosity of 36 (3000) fb$^{-1}$, corresponding the a value of $100 (39)$ using the latest ATLAS public analysis, are also marked.
}
\centering
\label{fig:upper_limit}
\end{figure}

\begin{figure}
\centering
\includegraphics[width=.70\textwidth]{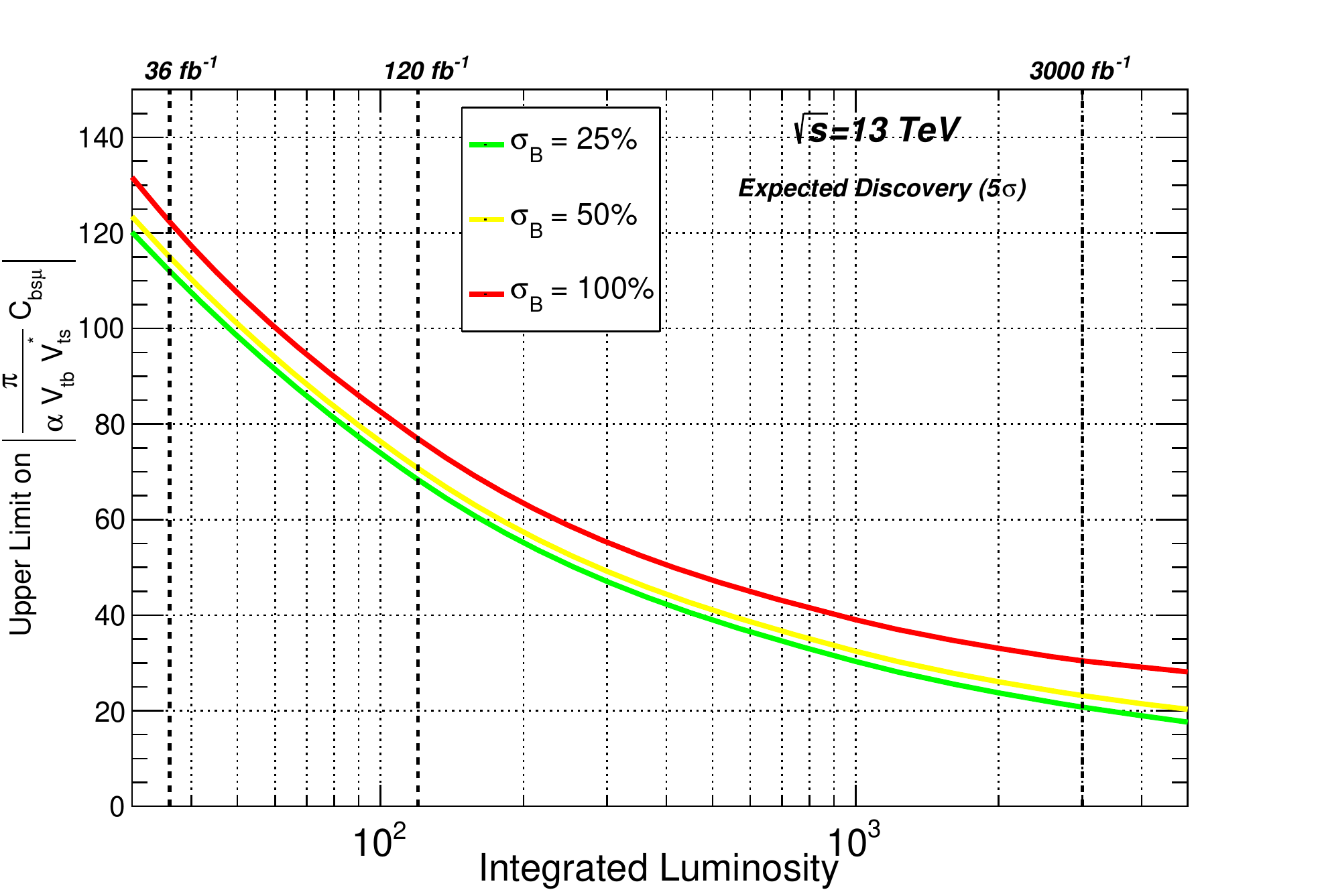}

\caption{Predicted discovery ($5\sigma$) on the term presented at equation \ref{eq:Cbslimit}, as a function of the total integrated luminosity.
The dashed lines correspond to integrated luminosity of (36, 120, 3000) fb$^{-1}$.
}
\centering
\label{fig:discovery}
\end{figure}

In order to quantitatively evaluate the gained sensitivity of the presented analysis, a calculation of the expected upper limit was done.
%similar to the one presented at \cite{Greljo:2017vvb}.
Figure \ref{fig:upper_limit} presents the expected upper limit on the term presented in equation \ref{eq:Cbslimit}, as a function of the total integrated luminosity.
Figure \ref{fig:discovery} presents the expected value of the term presented in equation \ref{eq:Cbslimit} for a discovery, as a function of the total integrated luminosity.

%using the selections presented in table \ref{tab:bsll_mumu_selections}.
The upper limit was calculated using the $CL_{s}$~\cite{ALRead2002} method.
Data was randomly chosen as an integer of a poisson distribution, with an average of the expected background yield.
The background only (signal+background) hypothesis was randomly chosen from a gaussian distribution with a mean of the background (signal+background) yield. 
%and standard deviation of $50\%$ of the background yield. 
Small differences were observed between different standard deviation assumptions on the background yield, so only the case of $50\%$ was considered. The uncertainty on the signal was not considered.
%This value originates from the number of standard deviation of a poisson distribution corresponding to a $p$-value of $0.05$ (95\% CL).
The expected upper limit for an integrated luminosity of (36, 120, 3000) fb$^{-1}$ is $\left| \frac{\pi}{\alpha V_{t b} V^*_{t s}} C_{b s\mu} \right| < (67, 36, 11)$, assuming 50\% of background uncertainty.
One can conclude that the dedicated analysis presented in this section improves the sensitivity significantly, compared to the limit presented in \cite{Greljo:2017vvb}.

\label{results}

\section{Conclusion}
Recent anomalies observed in the decays of B-mesons imply a new physics interaction that may lay between initial and final states involving a $b$-quark, a $s$-quark and a pair of opposite sign muons.
The interaction can be described in the scope of an EFT for a four-fermion interaction of a $b$-quark, an $s$-quark and a pair of muons. The four point interaction may play a role in the direct production of a $b$-quark and two opposite sign and same flavour leptons in $pp$ collisions.
We propose a search at the LHC for this type of interaction using final states with a pair of opposite sign muons and one $b$-jet. The search differs significantly from the analyses that target B-meson decays, mainly in the higher momentum thresholds of the objects involved in the final state.
When compared to previous ATLAS analysis, the search significantly improves the sensitivity to new physics laying in this four-fermion interaction.

%Specifically, we suggest the best up-to-date estimation of the sensitivity for this kind of search at the LHC. 
%We present the prospect for a search of new physics in this type of interactions at the LHC, in a process involves an s-quark, and a final state with two leptons and a b-jet.

\label{conclusion}

\acknowledgments

%This is the most common positions for acknowledgments. A macro is
%available to maintain the same layout and spelling of the heading.

We thank Michael E. Peskin for the useful discussion that steered us towards the idea of this paper.
We thank Yotam Soreq, Yael Shadmi and Shaouly Bar-Shalom for their help with the theoretical part of the paper.
This research was supported by a grant from the Unites States-Israel Binational Science Foundation (BSF), Jerusalem, Israel, and by a grant from the Israel Science Foundation (ISF).
%\paragraph{Note added.} This is also a good position for notes added
%after the paper has been written.

% BIBLIOGRAPHY
% use BIBTEX if you want
%\bibliographystyle{JHEP}
%\bibliography{yourBIBfiles}


\end{document}